\documentclass{llncs}
\usepackage[numbers,sectionbib]{natbib}

\usepackage{listings}
\usepackage{lstautogobble}
\AtBeginDocument{}
\usepackage{amsmath}
\usepackage{mathabx}
\usepackage{booktabs}
\usepackage{tikz}
\usepackage{pgfplots}
\usepackage{xspace}
\usepackage{syntax}
\usepackage{csquotes}
\usepackage{array}

\newcommand{\prefs}{\mbox{SPREFQL}\xspace}

\newcommand{\prefer}[1]{\succ_{\scriptscriptstyle{#1}}}
\newcommand{\notprefer}[1]{\not\succ_{\scriptscriptstyle{#1}}}

\newcommand{\code}[1]{{\tt #1}}

\newcommand{\term}[1]{\textit{#1\/}}
\newcommand{\quotes}[1]{`#1'}
\newcommand{\qquotes}[1]{``#1''}
\newcommand{\qstring}[1]{\texttt{"#1"}}

\newcommand{\nonterminal}[1]{\synt{#1}}

\begin{document}

\title{An extension of SPARQL for expressing qualitative preferences}

\author{Antonis Troumpoukis\inst{1}\inst{2} \and Stasinos Konstantopoulos\inst{1} \and Angelos Charalambidis\inst{1}}

\institute{%
   Institute and Informatics and Telecommunications, NCSR \quotes{Demokritos}
\\ Aghia Paraskevi 15310, Athens, Greece
\\ \email{\{antru,konstant,acharal\}@iit.demokritos.gr}
\and
   Department of Informatics and Telecommunications, University of Athens, Greece
}

\maketitle

\begin{abstract}
  In this paper we present \prefs, an extension of the SPARQL language
  that allows appending a \qstring{PREFER} clause that expresses
  \quotes{soft} preferences over the query results obtained by the
  main body of the query. The extension does not add expressivity and
  any \prefs query can be transformed to an equivalent standard SPARQL
  query. However, clearly separating preferences from the
  \quotes{hard} patterns and filters in the \qstring{WHERE} clause
  gives queries where the intention of the client is more cleanly
  expressed, an advantage for both human readability and machine
  optimization. In the paper we formally define the syntax and the
  semantics of the extension and we also provide empirical evidence
  that optimizations specific to \prefs improve run-time efficiency by
  comparison to the usually applied optimizations on the equivalent
  standard SPARQL query.
\end{abstract}


\section{Introduction}
\label{sec:intro}

Preferences can be used in situations where, while looking for the best
solution with respect to a set of criteria, we find out that too strict
criteria might not return any solutions, but relaxing them returns too
many solutions to sift through. The integration of preferences allows
to view some constraints as soft constraints that can be violated in
the former case and return less-preferred results, but will be
enforced in the latter case to only return more-preferred results.

Preferences have been explored in Artificial Intelligence~\cite{DHKP11},
Database Systems~\cite{StefanidisKP11}, Programming Languages~\cite{DelgrandeSTW04},
and, more recently, enjoy a growing interest in the area of the
Semantic Web~\cite{PivertST16}. In the Semantic Web context, preferences
allow users to sift through data of varying trustworthiness, quality,
and relevance from a specific end user's point of
view~\cite{dellavalle-etal:2013}. As argued by
\citet{SiberskiPT06}, the motivating example in the beginning of the
seminal Semantic Web article~\cite{berners2001semantic} can be
interpreted as a preference search.

Strictly speaking, preferences are not more expressive than standard
SPARQL. Their most prominent feature, returning less-preferred binding
sets in the absence of more-preferred ones, can be simulated
with \qstring{NOT~EXISTS} and, in general, with the syntax already
offered by SPARQL. However, clearly separating preferences from the
\quotes{hard} patterns and filters in the \qstring{WHERE} clause gives
us queries where the intention of the author is cleanly expressed and
not obscured. This has advantages in both human readability and
machine optimization.

In this paper, we first give a background on the treatment of
preferences in databases (Section~\ref{sec:bg})
and proceed to present our proposed \prefs syntax and semantics
(Section~\ref{sec:pref}). We then present our \prefs query processor
implementations and our benchmarks on them (Section~\ref{sec:exp}).
These empirical results are used to support our claim above
that optimizing directly at the \prefs syntax is more efficient than
rewriting into standard SPARQL and passing the latter to an
optimizing SPARQL query processor. We then present some related work
on the Semantic Web and compare it with our approach (Section~\ref{sec:disc}).
We close the paper with conclusions and future research directions
(Section~\ref{sec:conc}).

\section{Background}
\label{sec:bg}
\label{sec:bg:db}

Preference representation formalisms are either \term{quantitative},
where preferences are represented by a preference value
function~\cite{AgrawalW00,KoutrikaI04}, or \term{qualitative}, where
preferences are expressed by directly defining a binary preference
relation between objects~\cite{Chomicki03,Kiessling02}.
In the example below:
\begin{example}
\label{ex:qual-motivation}
Show me Sci-fi movies, assuming I prefer longer movies.
\end{example}
there is a hard constraint for SciFi movies and a preference towards
longer movies. Such a constraint can be represented both as a
quantitative function of the movies' runtime and as a qualitative
relation that compares movies' runtimes.
With this example, however:
\begin{example}
\label{ex:qual-motivation}
Show me Sci-fi movies, assuming I prefer original movies to their sequels.
\end{example}
it becomes apparent that there are cases where not all objects are
directly comparable, and therefore the total ordering implied by the
preference value function cannot always be defined.
In fact, \citet{Chomicki03} argues that the
qualitative approach is strictly more general than the quantitative
approach, as not all preference relations can be expressed using a
preference value function.
In Chomicki's framework, \term{preference relations} are
defined using first-order formulas:
\begin{definition}
  Given a relation schema $R(A_1,\dots,A_n)$ such that $U_i$, $1 \le i \le n$, is the
  domain of the attribute $A_i$, a relation $\prefer{}$ is a preference relation over 
  $R$ if it is a subset of $(U_1 \times\cdots\times U_n) \times (U_1 \times\cdots\times U_n)$. 
  A result tuple $t_1$ is said to be \emph{dominated} by $t_2$, if $t_2 \prefer{} t_1$.
\end{definition}
This general preference relation is restricted into
\term{intrinsic preference formulas} that do not rely on external information
to compare two objects:
\begin{definition}
  Let $t_1,t_2$ denote tuples of a given database relation.
  A {\em preference formula} $P(t_1,t_2)$ is a first-order formula defining a
  preference relation $\prefer{P}$ in the standard sense, namely, $t_1 \prefer{P} t_2$
  iff $P(t_1,t_2)$ holds. An {\em intrinsic preference formula} is a preference
  formula that uses only built-in predicates (i.e. equality, inequality, arithmetic 
  comparison operations, and so on).
\end{definition}

\begin{table}[b]
  \centering
  \caption{A sample movies relation.}
  \label{tab:moviesrel}
  \renewcommand*{\arraystretch}{1.1}
  \begin{tabular}{cllcc}
    \toprule
      \multicolumn{1}{c}{ID} &
      \multicolumn{1}{c}{Title} &
      \multicolumn{1}{c}{Genre} &
      \multicolumn{1}{c}{Duration} &
      \multicolumn{1}{c}{Sequel} \\
    \cmidrule{1-5}
    $m_1$ & Star Wars Ep.IV: A New Hope             & Sci-fi & 121 & $m_2$ \\
    $m_2$ & Star Wars Ep.V: The Empire Strikes Back & Sci-fi & 124 & $m_3$ \\
    $m_3$ & Star Wars Ep.VI: Return of the Jedi     & Sci-fi & 130 & \\
    $m_4$ & Die Hard                                & Action & 131 & $m_5$ \\
    $m_5$ & Die Hard with a Vengeance               & Action & 128 & \\
    \bottomrule
  \end{tabular}
\end{table}

\begin{example}
\label{ex:dbexample}
  Consider the \code{movie(ID,Title,Genre,Duration)} relation shown in
  Table~\ref{tab:moviesrel}. Suppose that we have the following preference:
  \quotes{I prefer one {\tt movie} tuple over another iff their genre is the same and the
    first one runs longer}.
  The preference relation $\prefer{P}$ implied by the previous
  sentence can be defined using formula $P$:
  \[ (i,t,g,d) \prefer{P} (i',t',g',d') \equiv (g = g') \land (d > d'). \]
  Therefore, we prefer movie $m_3$ to $m_2$, movie $m_2$ to $m_1$,
  $m_3$ to $m_1$ and movie $m_4$ to $m_5$. Both conjuncts must be
  satisfied for the preference relation to hold, so there is no
  preference relation between movies from different genres regardless
  of their runtime.
\end{example}

A new relational algebra operator is introduced, called {\em winnow}.
This operator takes two parameters, a database relation and a
preference formula and selects from its argument relation the most
preferred tuples according to the given preference relation.

Preference relations can be composed in order to form more complex
ones. Since preference relations are defined through preference formulas, 
in order to combine two such relations one must combine their corresponding formulas.
Given two preference relations $\prefer{P},\prefer{Q}$, the most common composition
operations are the following:

\begin{itemize}
\item {\em Boolean:} (e.g. intersection) 
      $t_1 \prefer{P \land Q} t_2 \equiv 
	(t_1 \prefer{P} t_2) \land (t_1 \prefer{Q} t_2),$
\item {\em Pareto:} 
      $t_1\prefer{P \otimes Q} t_2 \equiv 
	((t_1\prefer{P} t_2) \land (t_2 \notprefer{Q} t_1)) \lor 
	((t_1\prefer{Q} t_2) \land (t_2 \notprefer{P} t_1)),$
\item {\em Prioritized:} 
      $t_1 \prefer{P \triangleright Q} t_2 \equiv 
	(t_1 \prefer{P} t_2) \lor ((t_1 \sim_{\scriptscriptstyle{P}} t_2) \land (t_1 \prefer{Q} t_2)),$      
\end{itemize}
where $t_1 \notprefer{P} t_2 \equiv \neg (t_1 \prefer{P} t_2)$ and 
$t_1 \sim_{\scriptscriptstyle{P}} t_2 \equiv (t_1 \notprefer{P} t_2) \land (t_2 \notprefer{P} t_1)$.
      
In order to select the \quotes{best} tuples from a given relation $r$
based on a preference formula $P$, the {\em winnow} operator is
introduced:
\begin{definition}
  \label{def:winnow}
  Let $r$ be a relation and let $P$ be a preference formula defining a preference relation
  $\prefer{P}$. The {\em winnow} operator is defined as
  $$w_P\left(r\right) = \{ t \in r : \neg \exists t' \in r \mbox{ such that } t' \prefer{P} t \}.$$
\end{definition}

\begin{example}
  Given the relation {\tt movie} in Table~\ref{tab:moviesrel}
  and the preference formula $C$ of
  Example~\ref{ex:dbexample}, the result of the $w_P(\mathtt{movie})$
  operation is the movies with IDs $m_3$ and $m_4$.  $m_1$ and $m_2$
  are not included in the result because they are less preferred than
  $m_3$ and $m_5$ because it is less preferred than $m_4$. Since there
  is no preference relation between $m_3$ and $m_4$, they are both
  included in the result.
\end{example}

Although winnow can be expressed using standard relational algebra
operators~\cite{Chomicki03}, there also exist algorithms that directly
compute the result of the winnow operator $w_P(R)$. The most prominent such
algorithms are the \term{Nested Loops (NL)} algorithm and the
\term{Blocked Nested Loops (BNL)} algorithm.
In NL, each tuple of $R$ is compared with all tuples in $R$, therefore the complexity of NL is 
quadratic in the size of $R$. 
In BNL, a fixed amount of main memory (a {\em window}) is used, in order to keep a set of
incomparable tuples, which at the end of the algorithm will become the dominating tuples of $R$.
Even though the asymptotic time complexity of BNL is also quadratic, in practice BNL performs
better than NL. Especially in the case that the result set of winnow fits into the window,
the algorithm operates in one or two iterations (i.e. linear time to the size of $R$)~\cite{BorzsonyiKS01}.
Regarding the correctness of the result of each algorithm, NL produces the correct result 
for every preference relation (even in unintuitive cases such as preference relations in 
which a tuple is preferred to itself). On the other hand, BNL produces the correct result
only if the preference relation $\prefer{}$ is a
\term{strict partial order}~\cite{Chomicki03}, that is to say iff the relation is
(1) \term{irreflexive} $\neg( x \prefer{} x )$ 
(2) \term{transitive} $(x \prefer{} y) \land (x \prefer{} z) \Rightarrow (x \prefer{} z)$
and (3) \term{asymmetric} $(x \prefer{} y) \Rightarrow \neg(y \prefer{} x)$.

\begin{example}
  Let us assume the relation {\tt movie} in Table~\ref{tab:moviesrel}
  and the following preference formula $C'$:
  \begin{displayquote}
    \quotes{I prefer one {\tt movie} tuple over another iff their
    genre is the same and the first one has the second as sequel.}
  \end{displayquote}
  In this case, BNL is \emph{not} guaranteed to produce the correct result because
  $m_1$ \quotes{sequel} $m_2$ and
  $m_2$ \quotes{sequel} $m_3$, but
  $m_1$ \quotes{sequel} $m_3$ is not asserted, making
  the \quotes{sequel} property (and thus the whole preference
  relation) not transitive. The result of the BNL algorithm depends on
  the order in which pairs are tested: if $m_2$ is compared to $m_1$
  before being compared to $m_3$, the first comparison will remove
  $m_2$ from the window making $m_1$ and $m_3$ incomparable and the
  result is $\left\{m_1, m_3, m_4\right\}$; if $m_2$ is compared to
  $m_3$ before being compared to $m_1$, then both $m_3$ and $m_2$ will
  be removed and the result is $\left\{m_1, m_4\right\}$.
\end{example}

\section{The \prefs Language}
\label{sec:pref}

In this section we introduce \prefs, which is an extension of SPARQL
that supports the expression of qualitative preferences. User
preferences are expressed as a new solution modifier which eliminates
the solutions that are dominated by (i.e., are less preferred than)
another solution. This modifier is similar to a preference formula in
Chomicki's framework discussed above. In this section
we present the syntax and the semantics of \prefs, discuss its
expressive power, and we will give some examples of
\prefs queries.

\subsection{Syntax}

\begin{figure}[b!]
  \caption{The \prefs grammar.}
  \label{fig:syntax}

  {\it The full EBNF grammar for \prefs is the result of starting
    with the grammar for SPARQL~1.1 \cite[Section~19.8]{w3c-sparql},
    replacing Rule~18 with the first rule below, and appending the
    rest of the rules below.}

  \begin{grammar}
    <SolutionModifier> ::= [ <GroupClause> ] [ <HavingClause> ] [ <PreferClause> ]
    \\[0pt] [ <OrderClause> ] [ <LimitOffsetClauses> ] 
			
    <PreferClause> ::= `PREFER' <VarList> `TO' <VarList> `IF' <ParetoPref>
    
    <VarList> ::=  <Var> \alt `(' <Var>+ `)'
    
    <ParetoPref> ::=  <PrioritizedPref> [ `AND' <ParetoPref> ]
    
    <PrioritizedPref> ::=  <BasicPref> [ `PRIOR' `TO' <PrioritizedPref> ]
    
    <BasicPref> ::=  `(' <ParetoPref> `)' \alt <SimplePref>

    <SimplePref> ::=  <Constraint>

  \end{grammar}
\end{figure}

We assume as a basis the EBNF grammar that defines SPARQL syntax
\cite[Section~19.8]{w3c-sparql} and we extend it by changing the
definition of the \nonterminal{SolutionModifier} non-terminal
(Rule~18). The new definition adds a \nonterminal{PreferClause}
non-terminal between the \nonterminal{HavingClause} and the
\nonterminal{OrderClause} non-terminals. The rationale for this
positioning is that:
\begin{itemize}
  \item The prefer clause should be after the group-by/having clauses,
    as it would make sense to use in the former the aggregates
    computed by the latter.
  \item The prefer clause should be before the limit/offset clauses,
    as it would be counter-intuitive to miss preferred solutions
    because they have been limited out, so the limit should apply
    to the preferred solutions.
  \item The prefer clause could equivalently be either before or after
    the order-by clause, but there is no reason to sort solutions that
    are going to be discarded afterwards. Naturally an optimizer could
    also re-order these computations, but there is no reason why the
    default execution plan should not put these in the more efficient
    order already. A further advantage of placing the prefer clause
    before the order-by clause is that this avoids requiring from
    compliant \prefs implementations that they maintain the order of
    the result set.
\end{itemize}
Figure~\ref{fig:syntax} gives the EBNF rules that define
\nonterminal{PreferClause} and also re-define 
\nonterminal{SolutionModifier}. All non-terminals that are not defined
in this table are defined by standard SPARQL syntax:
\nonterminal{GroupClause} (Rule~19),
\nonterminal{HavingClause} (Rule~21),
\nonterminal{OrderClause} (Rule~23),
\nonterminal{LimitOffsetClauses} (Rule~25).
\nonterminal{Constraint} (Rule~69), and
\nonterminal{Var} (Rule~108).
Note, in particular, how basic preferences are a conjunction of
the standard SPARQL \nonterminal{Constraint} used in the definitions
of \qstring{HAVING} and \qstring{FILTER} clauses. This means that
preferences are expressed using the familiar syntax of SPARQL constraints.

In the remainder, we shall call \term{query base}
$B(\mathcal{Q})$ the standard SPARQL query that is derived from a
\prefs query $\mathcal{Q}$ by removing the \qstring{PREFER} clause.
We shall also call \term{full result set} the result set of
$B(\mathcal{Q})$ and \term{preferred result set} the result set of
$\mathcal{Q}$.
We continue with a simple example in \prefs.
\begin{example}
Suppose that we want to query an RDF database with movies and we have the
following preference:
\begin{displayquote}
  \quotes{I prefer one movie to another iff their genre are the same and the
  first one runs longer.}
\end{displayquote}
The size of the preferred result set is equal to the number of the available genres
in the dataset (since two films with different genre are incomparable). For
each genre, the selected film must be the one with the longest runtime.
The corresponding \prefs query is listed in Listing~\ref{lst:q1}.

To express preference of one binding set over another, we first use
the \qstring{PREFER} clause to assign variable names to the bindings
in the two binding sets, so that the two binding sets can be
distinguished from each other. We then use the \qstring{IF} clause to
express the conditions that make the first binding set dominate the
second one. In the query in Listing~\ref{lst:q1}, for example, there
are three bindings in each result, \texttt{(?title ?genre ?runtime)}.
In order to compare two binding sets, the \qstring{PREFER} clause
assigns the bindings in the first result to the variables 
\texttt{(?title1 ?genre1 ?runtime1)} and the bindings in the second
result to the variables \texttt{(?title2 ?genre2 ?runtime2)}.
These new variable names are then used in the \qstring{IF} clause to
specify when the first result dominates the second result. Notice that
any name can be used for the variables in the \qstring{PREFER} clause,
and what maps them to the variables in the \qstring{SELECT} clause is
the order of appearance.
For example, in this query, variables 
\texttt{?title1}, \texttt{?title2} correspond to variable \texttt{?title}, the variables 
\texttt{?genre1}, \texttt{?genre2} correspond to variable \texttt{?genre} and so on.
Note also that the names in the \qstring{PREFER} clause need to be
distinct from each other, but they do \emph{not} need to be distinct
from the names in the \qstring{SELECT} clause. In this manner, the
style shown in Listing~\ref{lst:alt-q1} is also possible, if the query
author prefers it.
\end{example}
Given the above, we define well-formed \prefs queries as follows:

\begin{definition}
  Let $\mathcal{Q} = \mathtt{SELECT} \; L \; \mathtt{WHERE} \; P_1 \;
    \mathtt{PREFER} \; L_1 \; \mathtt{TO} \; L_2 \; \mathtt{IF} \; P_2 \;$
  be a \prefs query produced by the grammar of Figure~\ref{fig:syntax}.
  Then, $\mathcal{Q}$ is \term{well-formed} iff $|L|=|L_1|=|L_2|$ 
  and all variables of $L_1,L_2$ are distinct.
\end{definition}

\begin{figure}[b!]

\begin{lstlisting}[
%  float=b,
  label=lst:q1,
  caption=\quotes{I prefer one movie over another iff their genre is
    the same and the duration of the first is longer}.,
]
SELECT ?title ?genre ?runtime WHERE {
 ?s a :film. ?s :title ?title. ?s :genre ?genre. ?s :runtime ?runtime.
}
PREFER (?title1 ?genre1 ?runtime1) TO (?title2 ?genre2 ?runtime2)
IF (?genre1 = ?genre2 && ?runtime1 > ?runtime2)
\end{lstlisting}

\begin{lstlisting}[
%  float=b,
  label=lst:alt-q1,
  caption=\quotes{I prefer one movie over another iff their genre is
    the same and the duration of the first is longer}.,
]
SELECT ?title ?genre ?runtime WHERE {
 ?s a :film. ?s :title ?title. ?s :genre ?genre. ?s :runtime ?runtime.
}
PREFER (?t ?genre ?runtime) TO (?otherT ?otherGenre ?otherRuntime)
IF (?genre = ?otherGenre && ?runtime > ?otherRuntime)
\end{lstlisting}

\begin{lstlisting}[
  label=lst:q2,
  caption=\quotes{Given two action movies, I prefer
    the longest one and more recent one 
    with equal importance}.,
]
SELECT ?title ?genre ?runtime WHERE {
 ?s a :film. ?s :genre :action. 
 ?s :title ?title. ?s :runtime ?runtime. ?s :year ?year.
}
PREFER (?title1 ?runtime1 ?year1) TO (?title2 ?runtime2 ?year2)
IF (?runtime1 > ?runtime2) AND (?year1 > ?year2)
\end{lstlisting}

\begin{lstlisting}[
  label=lst:q3,
  caption=\quotes{Given two action movies, I prefer
    the one that runs between
    115 and 125 minutes. If they are the same
    to me according to this criterion, 
    I prefer the ones that they are after 2005}.
]
SELECT ?title ?genre ?runtime WHERE {
 ?s a :film. ?s :genre :action. 
 ?s :title ?title. ?s :runtime ?runtime. ?s :year ?year.
}
PREFER (?title1 ?run1 ?year1) TO (?title2 ?run2 ?year2)
IF ( ?run1 >= 115 && ?run1 <= 125 && (?run2 < 115 || ?run2 > 125) ) 
PRIOR TO (?year1 >= 2005 && ?year2 < 2005)
\end{lstlisting}

\begin{lstlisting}[
  label=lst:q4,
  caption=\quotes{I want to watch a movie with
    \qquotes{Mad Max} in the title, and I prefer
    original movies to their sequels}.
]
SELECT ?film ?title WHERE {
 ?film a :film . ?film :title ?title. FILTER regex(?title, "Mad Max").
}
PREFER (?film1 ?title1) TO (?film2 ?title2)
IF EXISTS { ?film1 :sequel ?film2 }
\end{lstlisting}

\end{figure}

In Section~\ref{sec:bg} we presented some ways so that two preference relations can
be combined into one more complex one. As in the framework of Chomicki, we can
also use boolean operators to combine the individual boolean expressions 
(boolean composition). Besides logical operators, we offer the following
two preference combinators for combining preference relations:
\begin{itemize}
  \item Pareto composition: the \qstring{AND} combinator composes a
    relation from two preference relations that are of equal
    importance (cf. Listing~\ref{lst:q2}). We follow previous work
    \citep{KiesslingK02,SiberskiPT06} in using \qstring{AND} for
    the Pareto combinator, noting that it should not be
    confused with the logical conjunction operator.
  \item Prioritized composition: the \qstring{PRIOR TO} combinator
    composes a preference relation where the less-important right-hand
    side argument is only applied if the more-important left-hand side
    argument does not impose any preference between two object (cf.
    Listing~\ref{lst:q3}).
\end{itemize}

These combinations can be expressed within a simple constraint
with the elaborate use of boolean operators. But this
\quotes{syntactic sugar} makes useful expressions a lot more readable.
Compare, for example, the queries in Listings~\ref{lst:q2} and
\ref{lst:q3} with their equivalent queries without using the
\qstring{AND} and \qstring{PRIOR TO} combinators, in
Listings~\ref{lst:rewrite2} and \ref{lst:rewrite3} respectively.
 
\begin{figure}[t]

\begin{lstlisting}[
  label=lst:rewrite2,
  caption=Rewrite of the \qstring{PREFER} clause in Listing~\ref{lst:q2} without using the \qstring{AND} combinator.
]
PREFER (?title1 ?runtime1 ?year1) TO (?title2 ?runtime2 ?year2)
IF ( ((?runtime1 > ?runtime2) && !(?year2 > ?year1)) 
  || ((?year1 > ?year2) && !(?runtime2 > ?runtime1)) )
\end{lstlisting}

\begin{lstlisting}[
  label=lst:rewrite3,
  caption=Rewrite of the \qstring{PREFER} clause in Listing~\ref{lst:q3} without using the \qstring{PRIOR TO} combinator.
]
PREFER (?title1 ?run1 ?year1) TO (?title2 ?run2 ?year2)
IF ( (?run1 >= 115 && ?run1 <= 125 &&  (?run2 < 115 || ?run2 > 125)) 
     ||
     ( !(?run1 >= 115 && ?run1 <= 125 && (?run2 < 115 || ?run2 > 125)) &&
       !(?run2 >= 115 && ?run2 <= 125 && (?run1 < 115 || ?run1 > 125)) &&
        (?year1 >= 2005 && ?year2 < 2005) 
     ) )
\end{lstlisting}

\end{figure}

Since a basic simple preference is a \textit{Constraint}, anything that can appear
as a parameter in a SPARQL \qstring{FILTER} clause can be used as a simple basic
user preference, and has the same meaning as in SPARQL \qstring{FILTER} clauses.
This could be also an \qstring{EXISTS} expression, as it is shown in Listing~\ref{lst:q4}.
These type of preference relations are known as \term{extrinsic} preferences~\cite{Chomicki03},
and are not supported by Chomicki's framework. A preference relation is
extrinsic if the decision of whether an element is preferred over another depends 
not only on the values of the elements themselves, but also on external factors 
(such as the the \texttt{:sequel} predicate in our example).

\subsection{Semantics}

In this section we will define the semantics of \prefs. Our semantics extend the standard semantics of
SPARQL~\citep{w3c-sparql}. We assume basic familiarity of the semantics of SPARQL, but we will present
some basic terminology when needed.

We denote by $\mathbf{T}$ the set of all \term{RDF terms} and by $\mathbf{V}$ the set of all \term{variables}.
A \term{mapping} $\mu$ is a partial function $\mu : \mathbf{V} \to \mathbf{T}$.
The \term{domain} of a mapping $\mu$, denoted as $\mathsf{dom}(\mu)$ is the subset of $\mathbf{V}$ where
$\mu$ is defined.
It is straightforward to see that mappings express variable bindings
and that given a mapping $\mu$ it is always possible to construct a
\qstring{VALUES} clause that expresses the same bindings as $\mu$
does.

\begin{example}
\label{ex:mapping}
Let  $\mu =\{(\mathtt{g},\texttt{"Sci-fi"}), (\mathtt{r},121)\}$
Then $\mu$ expresses the same binding of variable \qstring{?g} as the
clause \verb|"VALUES ( ?g ?r ) { "Sci-fi" 121 }"|.
\end{example}

Following \citet{PerezAG09} we denote by $\ldbrack\cdot\rdbrack_D$ 
the \term{evaluation} of a SPARQL query over a dataset $D$.
If a query $Q$ is a SELECT query, then $\ldbrack Q\rdbrack_D$ is a set of mappings, which are the
solutions that satisfy $Q$ over $D$. If $Q$ is an ASK query, then $\ldbrack Q\rdbrack_D$ is equal
to $\mathsf{true}$ if there exists any solution for $Q$ in $D$, otherwise it is equal to $\mathsf{false}$.

We will now continue with the semantics of the preference solution modifier. Firstly though,
we have to include some preliminary definitions:

\begin{definition}
  Let $L=(l_1,\dots,l_n),B=(b_1,\dots,b_n)$  be two variable lists and 
  $\mu$ be a mapping s.t. $\mathsf{dom}(\mu) = \mathcal{B}$,
  where $\mathcal{B}$ is the set of all variables of $B$.
  Then, we denote by $\mathsf{Rename}_{B\to L}(\mu)$ 
  a mapping that is created from $\mu$ by renaming variable $b_i$ to $l_i$, 
  for all $i=1,\dots,n$.
\end{definition}

\begin{definition}
  Let $L,L',B$  be three variable lists, s.t. $|L|=|L'|=|B|$ and all variables that appear 
  in $L,L'$ are distinct. 
  Also, let $\mu,\mu'$ be two mappings s.t. $\mathsf{dom}(\mu) = \mathsf{dom}(\mu') = \mathcal{B}$,
  where $\mathcal{B}$ is the set of all variables of $B$.
  Then, we denote by $\mathsf{ConstructMapping}_{B\to L,B\to L'}(\mu,\mu')$
  a mapping such that
  $$\mathsf{ConstructMapping}_{B\to L,B\to L'}(\mu,\mu') = 
       \mathsf{Rename}_{B\to L}(\mu) \cup \mathsf{Rename}_{B\to L'}(\mu').$$
\end{definition}

\begin{definition}
  Let $C$ be a \emph{SPARQL Constraint} and $\mu$ be a mapping.
  Then, we denote by $\mathsf{ConstructQuery}(C,\mu)$ a query of the form
  \verb|"ASK { FILTER| C S \verb|}"|
  where $s$ is the \emph{SPARQL ValuesClause} that corresponds to the mapping $\mu$.
  Note: \emph{SPARQL Constraint} and \emph{SPARQL ValuesClause} as
  defined in the SPARQL specification \citep{w3c-sparql}.
\end{definition}

\begin{example}
\label{ex:semantics1}
Let  $\mu =\{(\mathtt{g},\texttt{"Sci-fi"}), (\mathtt{r},121)\}$, 
     $\mu'=\{(\mathtt{g},\texttt{"Sci-fi"}), (\mathtt{r},124)\}$,
$B =(\mathtt{g}, \mathtt{r} )$, 
$L =(\mathtt{g1},\mathtt{r1})$,
$L'=(\mathtt{g2},\mathtt{r2})$ and
$C=\qstring{( g1 = g2 \&\& r1 > r2 )}$.
Then, 
$$\mathsf{ConstructMapping}_{B\to L,B\to L'}(\mu,\mu') = \mu^* = 
  \left\{\begin{array}{c}
    (\mathtt{g1},\texttt{"Sci-fi"}), (\mathtt{r1},121), \\ 
    (\mathtt{g2},\texttt{"Sci-fi"}), (\mathtt{r2},124)
  \end{array}\right\}
$$
$$\mathsf{ConstructQuery}(C,\mu^*) = 
  \begin{array}{l}
      {\small \verb|"ASK { FILTER ( ?g1 = ?g2 && ?r1 > ?r2 )| }\\
      {\small \verb|       VALUES ( ?g1 ?r1 ?g2 ?r2 ) | }\\ 
      {\small \verb|              { ( "Sci-fi" 121 "Sci-fi" 124 ) } }"| }\\
  \end{array}
$$
\end{example}

As stated earlier, our preference solution modifier expresses a
preference relation between the results of the query base, therefore the
meaning of the \qstring{PREFER} clause is actually a binary predicate $p$ such that
$p(\mu,\mu')$ holds if $\mu$ is preferred over $\mu'$. Hence, below,
the \term{evaluation} $\ldbrack\cdot\rdbrack_D$ of a \qstring{PREFER} clause takes two mappings
as input. Recall that except from a simple \term{Constraint}, a preference relation can be 
expressed using the \term{Pareto} and \term{Prioritized} preference
compositors.\footnote{We use a slightly different notation in the
  following definitions from the definitions in Section~\ref{sec:bg}. Instead of
  writing $\ldbrack \mu\prefer{C}\mu'\rdbrack_{D,B}$ we write
  $\ldbrack{C}\rdbrack_{D,B}(\mu,\mu')=\mathsf{true}$. In addition,
  the operators $\triangleright$ and $\otimes$ correspond to the
  Prioritized and Pareto compositions.}

\begin{definition}
Let $D$ be a dataset.
Also, let $C$ be a constraint and $L,L',B$ be three variable lists,
s.t. $|L|=|L'|=|B|$ and all variables that appear in $L,L'$ are
distinct.
Also, let $\mu,\mu'$ be two mappings s.t.
$\mathsf{dom}(\mu) = \mathsf{dom}(\mu') = \mathcal{B}$,
where $\mathcal{B}$ is the set of all variables of $B$.
Then,
\begin{multline}\nonumber
  \ldbrack \mathtt{PREFER} \; L \; \mathtt{TO} \; L' \; \mathtt{IF}\; C \rdbrack_{D,B} =
  \left\{ \left(\mu,\mu'\right) : 
  \ldbrack \mathsf{ConstructQuery}(C,\mu^*) \rdbrack_D = \mathsf{true},\right.\\
  \left. \mu^* = \mathsf{ConstructMapping}_{B\to L,B\to L'}(\mu,\mu')
  \right\}
\end{multline}
Composite clauses using the \qstring{PRIOR TO} and \qstring{AND}
combinators are defined as follows:
\begin{enumerate}
         \item $\ldbrack \mathtt{PREFER} \; L \; \mathtt{TO} \; L' \; \mathtt{IF}\; {P}\;\mathtt{PRIOR}\;\mathtt{TO}\;Q \rdbrack_{D,B} = 
	   \ldbrack \mathcal{P} \rdbrack_{D,B} \triangleright \ldbrack \mathcal{Q} \rdbrack_{D,B}$,
	   
    \item $\ldbrack \mathtt{PREFER} \; L \; \mathtt{TO} \; L' \; \mathtt{IF}\; {P}\;\mathtt{AND}\;Q \rdbrack_{D,B} = 
	   \ldbrack{P}\rdbrack_{D,B} \otimes \ldbrack{Q}\rdbrack_{D,B}$,
\end{enumerate}
where
$\mathcal{P} = \mathtt{PREFER} \; L \; \mathtt{TO} \; L' \; \mathtt{IF}\; {P}$,
$\mathcal{Q} = \mathtt{PREFER} \; L \; \mathtt{TO} \; L' \; \mathtt{IF}\; {Q}$,
$C$ is a constraint expression and $P,Q$ non-terminal symbols.
\end{definition}

Notice that in Example~\ref{ex:semantics1}, 
$\ldbrack \mathtt{PREFER} \; L \; \mathtt{TO} \; L' \; \mathtt{IF}\; C \rdbrack_D (\mu,\mu') = \mathsf{true}$
for every dataset $D$, or in other words the evaluation of the corresponding preference predicate is independent
from the dataset $D$. This is the case for all constraint expressions that use only built-ins. The reason why we use the construction of this ASK query, 
is in the case of preferences that are defined with the use of an EXISTS expression (see for example Listing~\ref{lst:q4}).
In that example, in order to check whether a mapping is preferred from another, 
one has to check the dataset $D$ for the existence of the corresponding \texttt{:sequel} triple.

Having defined the meaning of preference relations, we can proceed to
define how the preference solution modifier uses a preference relation
to reduce the full result set of the query base into the preferred
result set. For this,
we refer to the \term{winnow} operator $w_P(\ldbrack{Q}\rdbrack_{D})$
which outputs the preferred result set
when given the preference relation $P$ and the full result set
$\ldbrack{Q}\rdbrack_{D}$ (cf. Definition~\ref{def:winnow}).

\begin{definition}
  Let $Q$ be a SELECT query. Then, we denote by
  $\mathsf{ProjVarList}(Q)$ the projection list
  in the same order that it appears in the SELECT clause.
\end{definition}

\begin{definition}
  Let $D$ be a dataset, $Q$ be a SELECT query and $L,L'$ be two variable lists
  such that $|\mathsf{ProjVarList}(Q)|=|L|=|L'|$ and all variables that appear in $L,L'$ are distinct. 
  Then, 
  $$\ldbrack Q \; \mathtt{PREFER} \; L \; \mathtt{TO} \; L' \; \mathtt{IF}\; C \rdbrack_D = 
      w_{\ldbrack \mathtt{PREFER} \; L \; \mathtt{TO} \; L' \; \mathtt{IF}\; C \rdbrack_{D,B}} (\ldbrack{Q}\rdbrack_{D}),$$
  where $B=\mathsf{ProjVarList}(Q)$ and $C$ be a non terminal symbol.
\end{definition}

\subsection{Expressive power of \prefs}

Winnow can be expressed using standard relational algebra operators~\cite{Chomicki03}. 
Therefore, a \prefs query, which is essentially a SPARQL 1.1 query extended with a winnow operation, can
be also expressed using standard SPARQL 1.1, using a \qstring{NOT EXISTS} query rewriting.
Given a \prefs query of the form
$$\mathtt{SELECT}\;{L}\;\mathtt{WHERE}\;\{\;{P}\;\}\;
  \mathtt{PREFER}\;{L_1}\;\mathtt{TO}\;{L_2}\;\mathtt{IF}\;{C}\;$$
the preferred result set consists of the result mappings of the query base that are the
most preferred ones, or equivalently all mappings in the full result set such that there does not exist 
any mapping that is more preferred. This fact can be expressed using a standard SPARQL query of
the following form
$$\mathtt{SELECT}\;{L}\;\mathtt{WHERE}\;\{\;{P}\;
  \mathtt{FILTER}\;\mathtt{NOT}\;\mathtt{EXISTS}\;\{\;
  {P_{\{L/L_1\}}}\;\mathtt{FILTER}\;{C_{\{L_2/L\}}}\;\}\;\}$$
where $P_{\{L/L_1\}}$ is created by $P$ by replacing all variable names of $P$ that appear in $L$ with its
corresponding variable in $L_1$, and $C_{\{L_2/L\}}$ is created by $C$ by replacing all variable
names of $C$ that appear in $L_2$ with its $L$. The remaining variables on the new constructions are
replaced with fresh variables. If $C$ is a Pareto or a Prioritized composition, we first apply
the rewritings into their corresponding simple preferences (ref. Section~\ref{sec:bg},
Listing~\ref{lst:rewrite2} and Listing~\ref{lst:rewrite3}).
For example, the corresponding rewriting of Listing~\ref{lst:q1} is illustrated in Listing~\ref{lst:rewrite}.

\begin{lstlisting}[
  float=b,
  label=lst:rewrite,
  caption=Rewriting of Listing~\ref{lst:q1} into standard SPARQL.,
]
SELECT ?title ?genre ?runtime
WHERE {
  ?s a :film. ?s :title ?title. ?s :genre ?genre.
  ?s :runtime ?runtime.
  FILTER NOT EXISTS {
    ?s_tmp a :film. ?s_tmp :title ?title1. ?s_tmp :genre ?genre1. 
    ?s_tmp :runtime ?runtime1.
    FILTER (?genre1 = ?genre && ?runtime1 > ?runtime)  }
}
\end{lstlisting}

Comparing the two queries, we observe that the \prefs query is smaller
(it contains half the number of triple patterns), and it separates
the definition of preferences from the hard constraints. 
This separation alleviates the need for the query author to include in the 
query body the actual operation that performs the selection of the
best solutions, and to express the desired definition of preferences
is more clearly. Apart from the advantages in human readability, 
there exist advantages in machine optimization as well.
It would be difficult for a general purpose SPARQL optimizer to find
out that in the query in Listing~\ref{lst:rewrite} actually implements
an operation that resembles a self-join and the result can be computed
even in a single pass (as in BNL algorithm).

\section{Experiments}
\label{sec:exp}

\subsection{Implementation and experimental setup}

This section experimentally validates the idea that optimizations
specific to \prefs (such as efficient implementations of the winnow operator)
can improve the overall query performance in comparison to the equivalent standard
SPARQL query and its standard optimizations. As a proof of concept,
we provide an open source prototype implementation of 
\prefs.\footnote{\label{ft:code}cf. \url{https://bitbucket.org/dataengineering/sprefql}}
Our implementation is developed in Java within the RDF4J
framework,\footnote{cf. \url{http://rdf4j.org}}
and it includes two implementations of the winnow operator (i.e using NL and BNL
algorithms) and a query rewriter which transforms a \prefs query into the equivalent
SPARQL query, using the \qstring{NOT EXISTS} transformation. 
Our evaluator has the ability to operate over a simple memory store 
using the standard RDF4J evaluation mechanism, or over a remote SPARQL endpoint,
in which the query base is executed.

In this experiment we are performing \prefs queries on the 
LinkedMDB database.~\footnote{cf. \url{http://www.linkedmdb.org}}
Our query set contains 7 queries. The queries are:
{\bf Q1:} Listing~\ref{lst:q1},
{\bf Q2:} Listing~\ref{lst:q2},
{\bf Q3:} Listing~\ref{lst:q3},
{\bf Q4:} Listing~\ref{lst:q2} without genre restriction,
{\bf Q5:} Listing~\ref{lst:q3} without genre restriction,
{\bf Q6:} Listing~\ref{lst:q4} and
{\bf Q7:} Listing~\ref{lst:q4} without the FILTER, but for all movies
that feature the character \quotes{James Bond},
instead.\footnote{These listings are edited in the paper for
  conciseness. The exact queries used in the experiment can be found
  at our code repository, cf. Footnote~\ref{ft:code}.}
Firstly, we issue the query bases for each \prefs query directly on the SPARQL endpoint,
and then we evaluate all \prefs queries, using i) the NL algorithm, ii) the query rewriting
method and iii) the BNL algorithm.
The window size for the BNL algorithm was set large enough
to contain all results, since we know that BNL behaves better if the
preferred result set fits entirely in the window.
The experiment was performed on a  Linux machine (Ubuntu 14.04 LTS)
with a 4-core Intel(R) Xeon(R) CPU E31220 at 3.10GHz and 30 GB RAM.
The LinkedMDB dataset was loaded into a locally deployed Virtuoso
SPARQL endpoint.~\footnote{Community edition Version 7.1,
cf. \url{http://virtuoso.openlinksw.com}}

\subsection{Results}

Table~\ref{tab:resultsvirt} gives the experimental results.
We observe that NL has the worst query execution times, and its performance
is quadratic in the execution time of the query base. On the first 6 queries,
BNL performs better than rewriting. Since BNL was configured so that to perform
at its best, the query execution time of BNL is in most cases almost equal to
that of the query base. The difference between the execution times of BNL and
the query base in Q4 and Q5, can be explained due to the fact that the full
result set is larger and BNL has to make more comparisons to calculate the
preferred result. The rewrite method in those cases performs much worse than BNL
(but much better than NL). In Q7 though, where an extrinsic preference is
expressed, we have a different situation. The comparisons that BNL has to make
are not that many (they are at most $23 \cdot 22$), but here BNL has to consider
the database each time in order to decide whether one solution is preferred over
another. So, BNL issues a heavy load of ASK queries to the endpoint,
and therefore rewriting outperforms BNL in Q7. This also explains why
BNL has a comparable execution time for Q7 and Q1, although Q1
fetches and considers orders of magnitude more results than Q7.
As Q6 also expresses an extrinsic preference, we would expect
query rewriting to outperform BNL, but the base result set is
very small and the cost to prepare the rewrite is not recuperated.
Overall, in our experiments BNL performed better in
intrinsic preferences while rewriting performed better in extrinsic
preferences.

In the last two queries, we observe that the number of the results that BNL returns
is greater than the expected result. This happens because here the preference
relation (which is the same for Q6 and Q7) is not a transitive relation (the
\texttt{:sequel} is not a transitive predicate). This is a known issue of BNL,
since BNL returns the correct number of results only on preference relations that
impose a \term{strict partial order} (cf. Section~\ref{sec:bg:db}). Therefore,
in terms of the correctness of the result, rewriting is better than BNL for non
strict partial order intrinsic preferences (in extrinsic preferences, rewriting is
preferred anyway due to time performance). Checking whether an
intrinsic preference expression corresponds to a strict partial order
relation is not computationally challenging, as it depends only the
size of the expression itself~\citep[Section 3.1]{Chomicki03}. In
extrinsic expressions, transitivity needs to be confirmed extensionally by
issuing \qstring{ASK} queries.

\begin{table}[b!]
  \centering
  \caption{Number of returned results and query execution time
    (in milliseconds) for NL, query rewriting, and BNL. For BNL, the
    number of binding sets that need to be maintained in memory is
    also given, and the total number of bindings in these sets.}
  \label{tab:resultsvirt}
  \begin{tabular}{l *{4}{>{\bfseries}rr} rr}
    \toprule
      & \multicolumn{2}{c}{Query Base}
      & \multicolumn{2}{c}{NL}
      & \multicolumn{2}{c}{rewrite}
      & \multicolumn{4}{c}{BNL} \\
    \cmidrule(l){2-3}
    \cmidrule(l){4-5}
    \cmidrule(l){6-7}
    \cmidrule(l){8-11}
      & \multicolumn{1}{c}{exec.}
      & \multicolumn{1}{c}{num.} 
      & \multicolumn{1}{c}{exec.}
      & \multicolumn{1}{c}{num.} 
      & \multicolumn{1}{c}{exec.}
      & \multicolumn{1}{c}{num.} 
      & \multicolumn{1}{c}{exec.}
      & \multicolumn{1}{c}{num.}
      & \multicolumn{1}{c}{num.}
      & \multicolumn{1}{c}{num.} \\
      & \multicolumn{1}{c}{time}
      & \multicolumn{1}{c}{res.} 
      & \multicolumn{1}{c}{time}
      & \multicolumn{1}{c}{res.} 
      & \multicolumn{1}{c}{time}
      & \multicolumn{1}{c}{res.} 
      & \multicolumn{1}{c}{time}
      & \multicolumn{1}{c}{bindsets}
      & \multicolumn{1}{c}{bindings}
      & \multicolumn{1}{c}{res.} \\
    \midrule
    Q1 & 556 & 6,955 & 1,613,515 &  36 &   4,750 &  36 &   812 &  36 & 108 &  36 \\
    Q2 &  52 &   390 &     9,124 &   5 &     188 &   5 &    65 &   6 &  18 &   5 \\
    Q3 &  52 &   390 &    10,530 &   8 &     254 &   8 &    91 &   8 &  24 &   8 \\
    Q4 & 872 & 9,612 & 3,272,789 &   8 & 197,044 &   8 & 1,238 &   9 &  27 &   8 \\
    Q5 & 872 & 9,612 & 3,452,048 & 108 & 193,338 & 108 & 2,370 & 108 & 324 & 108 \\
    Q6 & 135 &     4 &       794 &   1 &     296 &   1 &   170 &   2 &   4 &   2 \\
    Q7 &  85 &    23 &     1,276 &   2 &      93 &   2 &   820 &   8 &  16 &   8 \\
    \bottomrule
  \end{tabular}
\end{table}

Regarding the memory footprint of the BNL algorithm, since BNL only
maintains the current set of undominated results it is expected to
require considerably less space than the base result set. In most
cases, the maximum number of results maintained in memory will be
close to the final number of results. In our experiments, only Q2 and
Q4 required a slight amount of extra space, which can happen when many
results that do not dominate each other are received before a result
that dominates them.

\section{Related Work}
\label{sec:disc}

In the Semantic Web literature there have been proposed SPARQL extensions
that feature the expression of preferences~\cite{PivertST16},
typically transferring ideas and results from
relational database frameworks much like the work presented here.

When it comes to quantitative preferences, prominent examples include
the extensions proposed by~\citet{ChengMY10}
and~\citet{magliacane-etal:2012}. Closer to our work, influential databases
research on \emph{qualitative} preferences includes the work
of Kie{\ss}ling~\cite{Kiessling02,KiesslingK02}. This was used by
\citet{SiberskiPT06} to propose a SPARQL extension
using a \qstring{PREFERRING} solution modifier.
Contrary to our approach, these preferences are expressed using unary
preference constructors. These constructors are of two types:
\term{boolean preferences} where the preferred elements fulfill a
specific boolean condition while the non-preferred do not; and 
\term{scoring preferences}, denoted with a \qstring{HIGHEST} or \qstring{LOWEST} keyword,
where the preferred elements have a higher (or lower) value from the 
non preferred ones. Simple preferences expressed with these
constructors can be further combined using Pareto and prioritized
composition operators. 
\citet{GueroussovaPM13} further extended this language with an
\qstring{IF-THEN-ELSE} clause which allows expressing
\term{conditional preferences} that apply only if a condition
holds. Conditional preferences allow several other
\quotes{syntactic sugar} preference constructors to be defined,
such as \qstring{AROUND} and \qstring{BETWEEN}.

By comparison, the work presented here is (to the best of our knowledge) the first one
to transfer to the Semantic Web the more general framework by
\citet{Chomicki03}, allowing the expression of extrinsic
preferences. Each of the basic preference constructors (boolean, scoring and conditional
preferences) as well as the compositions in the approaches by~\citet{SiberskiPT06}
and~\citet{GueroussovaPM13} can be transformed in \prefs.
For example, a query of the form 
\begin{lstlisting}
  SELECT ?s ?o WHERE {?s :p ?o} PREFERRING HIGHEST(?o)
\end{lstlisting}
can be transformed into \prefs:
\begin{lstlisting}
  SELECT ?s ?o WHERE {?s :p ?o} PREFER (?s1 ?o1) TO (?s2 ?o2) IF (?o1>?o2)
\end{lstlisting}
Since in \prefs the preference relation is expressed using a binary formula, 
the reverse translation is not always possible
(for example in Listings~\ref{lst:q1} and~\ref{lst:q4}).

\section{Conclusions and Future Work}
\label{sec:conc}

In this paper we propose \prefs, an extension of SPARQL that allows
the query author to specify a preference that modifies the query
solutions. Although a \prefs query can be transformed into standard
SPARQL, standard SPARQL query processing misses opportunities to
optimize execution by avoiding the exhaustive comparison of all
solution pairs. Our experiments demonstrate that when the BNL
algorithm is applicable, even for relatively small result sets of
under 10k tuples its execution can be two orders of magnitude faster
than that of state-of-art SPARQL query processors.

Our first future work direction will be to evaluate the mean gain that
can be achieved on realistic workflows. We plan to achieve this by
identifying potential test cases where the \prefs extensions can be
used, so that we can estimate how often the BNL optimization is
applicable. This will also help us further develop the language,
identifying additional \quotes{syntactic sugar} constructs that can
hint at optimizations targeting intransitive relations that fall
outside the scope of BNL. Further extensions could allow the client
to refer to preferences and preference-related metadata within the
knowledge base itself \cite{lukasiewicz-etal:2013,PoloMBRG14,RosatiNLLM15}.

A more ambitious future extension is to allow the client application
to not only request the most preferred results, but to also be able to
request all results ordered in different \quotes{layers} of
preference. This is a more general solution than any quantitative
preference ranking system, as it handles the full
generality of partially ordered preferences. We plan to base this on
graph-theoretic work in sequencing and scheduling, such as the
Coffman-Graham algorithm \citep{coffman-graham:1972} which is widely
used to visualize graphs as layers panning out
of a central vertex. By representing arbitrary (including
partial-order) preference relations as a directed graph, we can use
similar layering approaches to order results in such a way that no
dominated tuple is returned before any of the tuples that dominate it.

\section*{Acknowledgements}

The work described here has received funding from the European Union's
Horizon~2020 research and innovation programme under grant agreement
No~644564.
For more details, please visit
\url{https://www.big-data-europe.eu}

\bibliographystyle{splncsnat}

\end{document}